# DefectTwin: When LLM Meets Digital Twin for Railway Defect Inspection

Rahatara Ferdousi (School of Electrical Engineering and Computer Science, University of Ottawa, Ottawa, Ontario, Canada; Email: rferd068@uottawa.ca),
M. Anwar Hossain (School of Computing, Queen's University, Kingston, Ontario, Canada),
Chunsheng Yang (Institute of Artificial Intelligence, Guangzhou University, Guangzhou, China),
Abdulmotaleb El Saddik (School of Electrical Engineering and Computer Science, University of Ottawa, Ottawa, Ontario, Canada).

***Abstract*—A Digital Twin (DT) replicates objects, processes, or systems to enable real-time monitoring, simulation, and predictive maintenance. Recent advancements, such as Large Language Models (LLMs), have revolutionized traditional AI systems and show immense potential when combined with DT in various industrial applications. Railway defect inspection is one such application, which traditionally requires a large volume of defect samples to identify underlying patterns. However, training a new defect classifier with limited samples often leads to overfitting and poor performance on unseen defects. This challenge can be addressed by integrating pre-trained LLMs into DT, as specialized LLMs for defect inspection inherently reduce the need for extensive sample data. We propose an integration between LLM and DT for railway defect inspection and enable its usage in consumer electronics (CE) devices. Accordingly, we introduce DefectTwin, which utilizes a multimodal and multi-model (M²) LLM-based AI pipeline to analyze seen and unseen visual defects in railways. Using this application, a railway agent can mimic the tasks of an expert defect analyst using CE devices (e.g., tablets). The multimodal processor in DefectTwin ensures that the response generated from the AI pipeline is in a consumable format. An instant user feedback handling mechanism (instaUF) enables the Quality-of-Experience (QoE) feedback-loop within DefectTwin. The proposed M² LLM outperforms existing base models by achieving high precision (between 0.76-0.93) across multimodal input (text, image, video) that characterizes pre-trained defects. Additionally, we obtained better performance in zero-shot generalizability for unseen defects. We also evaluated the latency, token count, and usefulness of the responses generated by the DefectTwin application on a CE device. To the best of our knowledge, DefectTwin is the first LLM-integrated DT for railway defect inspection.**

***Index Terms*—Digital Twin, Large Models, Large Language Models, Consumer Electronics, Visual Railway Defect Inspection, Multimodal LLM, Multimodal AI**

## I. Introduction

DT – Artificial Intelligence (AI) integrated DTs deployed in Consumer electronics (CE) are beneficial for various industrial applications [1], [2]. For instance, applications like railway defect inspection have recently gained attention [3] in this area. However, existing AI-integrated systems often struggle with the complexity of visual inspection tasks [4]. Such complexity mainly arises from the limited defect samples, leading to suboptimal performance [5], [6].

Recent advancements like Large Language Models (LLMs), have revolutionized traditional AI systems by excluding the need for rapid training on huge amounts of samples. The characteristics of LLM to learn continuously from new data enhance the performance in unseen classification tasks (also known as zero-shot generalizability). This motivates us to improve the accuracy, efficiency, and generalizability of railway defect detection by combining DT and LLM. Therefore, in this research, we develop a case where LLM-integrated DT applications are designed for use in CE devices.

In this research, we introduce DefectTwin as an LLM-integrated DT system for visual railway defect inspection [1]. DefectTwin uses a synthetic dataset generation pipeline to create a custom visual instruct dataset, which fine-tunes a base language model (e.g., GPT-3.5) [7] into a specialized Defect LLM. The proposed system enhances user prompts by integrating Virtual Prompt Injections (VPI) [8] and system messages [9]. This enhanced prompt is processed by the fine-tuned LLM to generate a detailed visual defect description. Multimodal models use this description to create [10] or understand [11] images, videos, or 3D models of defects. An Integrated Multimodal Processor [12] ensures user-friendly output consumed on CE device. In addition, the Instant User Feedback (InstaUF) Pipeline iteratively improves the performance through user feedback.

The key research question investigates whether LLM-integrated DTs can enhance defect detection accuracy and maintenance efficiency in railway systems. The apps prototyped for DefectTwin are tested on a CE device (iPad 10th generation). The experiment involves synthetic dataset generation, fine-tuning, and validation using real-world datasets from the Canadian Pacific Railway (CPR) and other sources. We observed that DefectTwin achieves a precision of 0.93 in identifying railway defects, outperforming existing models.

Our research includes an ablation study addressing the impact of integrating text, image, and video data on defect detection accuracy, the effectiveness of synthetic datasets, the enhancement of model performance and user satisfaction through the QoE feedback loop, and the comparative performance of the proposed algorithms. Results confirm DefectTwin as an effective solution for automated visual rail defect inspection. Additionally, we proved the optimality of the proposed algorithms theoretically.

[1]Please find the codes and data used in this research

Key contributions of this paper include:

1) Introducing a Multimodal and Multimodel (M²) LLM-based AI inferencing Pipeline in DT for a specialized case- visual railway defect inspection.
2) Proposing a pipeline for generating synthetic datasets to address data scarcity, improving zero-shot generalizability in domain-specific LLM fine-tuning, which was not addressed in our previous work [13].
3) Designing Algorithm for Instant User Feedback Handling Loop for continuous model refinement.
4) Employing a multimodal processor to refine generative media to increase usefulness (e.g., enhanced defect analysis).

Subsequent sections cover the literature review in Section II, methodology in Section III, experimental setup and results in Section IV, discussion, and conclusion, detailing the development and validation of the DefectTwin system, its implications, and future research directions in section V.

## II. Literature Review

In this section, we synthesize the literature on integrating LLMs into DT systems for visual railway defect analysis [2].

### A. DT and AI

DT technology presents transformative opportunities by allowing manufacturers to simulate real-world conditions digitally, enhancing product design, development, and maintenance [14]–[16]. DTs enable virtual simulation of objects and processes to optimize risk and cost, offering personalized user experiences through continuous data collection and analysis from consumer devices [6].

Advancements in AI have led to the development of LLMs like ChatGPT and GPT-4, capable of addressing complex problems [17]. Integrating LLMs and Visual LLMs (VLMs) [18] into CE can revolutionize device design and usage, optimizing production processes and predicting equipment failures [6]. Enhanced user interaction and voice-activated assistant accuracy are key benefits [12], [19]. However, challenges such as sustainability, data privacy, and on-device LLMs must be addressed [7].

### B. Multimodal LLMs for AI-Integrated DT

As the demand for intelligent automation continues to grow, AI-integrated DT and LLMs are expected to play a pivotal role in the future of inspection in manufacturing and other applications [7] [5]. While the general impact of LLMs and DT has been discussed in the context of consumer electronics [5] [19] [20], evaluations of LLM-integrated DT in specialized domains remain limited.

In this research, we focus on a specific use case: visual railway defect inspection using a CE device. We use a tablet to explore and evaluate the DefectTwin apps. So that the impact of LLM-integrated DT in resource-constrained CE devices can be understood. The following sections detail various LLM methods we identified as applicable to our target use case, highlighting their potential in railway defect inspection through enhanced precision and efficiency.

*1) Text-to-Image Models:* While text-to-image models [21] offer a potential solution for generating synthetic data, creating effective prompts for specialized domains remains challenging [22]. Instruction tuning, which fine-tunes models to respond to specific prompts, can enhance the relevance and usefulness of generated images for specialized tasks [5].

*2) Hybrid Instruction-Following Agents:* There are two main approaches to building instruction-following agents [23]: (i) Multimodel, which coordinates various models via frameworks like LangChain or LLMs (e.g., Visual ChatGPT), and (ii) Multimodal, which can support multimodal input-output. A hybrid instruction-following agent that integrates both multimodal and multimodel approaches is ideal for comprehensive task handling [24].

*3) High-Quality Instructions:* High-quality instructions are crucial for fine-tuning multitasking agents. Recent trends involve using GPT-generated instructions (e.g., LLAVA [9], Objaverse [25], and MIMIC-IT [26]) to improve performance by helping models better understand the context and specific domain requirements. To ensure the quality of generated instruction-response pairs, the Syphus pipeline [26] incorporates system messages, visual annotations, and in-context examples as prompts for ChatGPT. This approach helps maintain high standards in instruction generation.

*4) Positioning Attacks:* Positioning attacks, where agents receive continuous prompts without relevant visual details, can degrade performance. The Virtual Prompt Injection (VPI) [8] pipeline addresses this issue using trigger instructions and virtual prompts to ensure the agent captures the correct context.

*5) Incorporating Human Feedback for Quality Assurance:* Hybrid reinforcement learning models that incorporate human feedback (HF) and AI feedback are essential for ensuring the quality of responses, especially in specialized domains like railway defect inspection [24].

### C. Requirements and Challenges

Key requirements and challenges in deploying LLM-integrated DT systems for visual defect inspection include:

- **Data Scarcity:** Rich synthetic data generation is essential because it helps overcome the limitations of real-world data availability, thereby improving model performance and increasing the diversity of training examples. This ensures that the models can generalize better to a wide variety of defect scenarios [13], [27].
- **Effective Prompts:** Instruction tuning is crucial as it enhances the relevance and usefulness of generated images for specialized tasks. By fine-tuning the prompts, LLMs can produce more accurate and task-specific outputs, which is particularly important in specialized domains like defect inspection [5].
- **Hybrid Instruction-Following Agents:** Multimodal and multi-model capabilities are vital because they combinedly improve the in-context response generation capacity of LLMs. This means that the LLMs can better understand and respond to complex inputs that include multiple data types (e.g., text, images, videos), resulting in more accurate and context-aware defect analysis and predictions.

[2]Please check the paper collection here.

## III. Methodology

This section first presents an overview of the DefectTwin framework followed by its components and algorithms.

The proposed DefectTwin system is an LLM-integrated DT approach designed to enhance railway defect detection, predictive maintenance, and user interaction. The system comprises several interconnected components, as illustrated in Fig. 1 [3]. As in Fig. 1, different sensing devices are employed to collect defect parameters such as images and metadata from the defective physical railway components. This data is transmitted to data storage and data pre-processing units, which are bidirectionally connected to the AI inferencing engine. This connection ensures high defect identification accuracy through data augmentation and synthetic data generation. Users interact with the system via a multimodal interface, utilizing different forms of DT (e.g. information-twin and predictive-twin) for defect analysis and maintenance. The information twin makes decisions by analyzing existing data, such as identifying the type of defect found. In contrast, the predictive twin anticipates future states in a simulative manner, like predicting additional cracks on the track before they occur. The system includes a Quality-of-Experience (QoE) feedback loop to continuously refine AI models based on user feedback.

The proposed $M^2$ LLM-based AI Inferencing Pipeline aims to generate a high-fidelity in-domain synthetic dataset for fine-tuning a base LLM to improve the performance of multimodal decoders used for various purposes, such as text-to-image, video-to-text, and image-to-text. The multimodal approach supports various data types, while the multimodel approach uses task-specific models (like video-to-text or text-to-image) accessed through a fine-tuned LLM. The pipeline involves several steps as illustrated in Fig. 2, including generating synthetic data, fine-tuning the base LLM, and integrating multimodal processing to dynamically map generated defect textures to 3D models of rail components. The key components in the proposed AI inferencing pipeline are described as follows.

### A. Synthetic Defect LLM Dataset Generation

In our proposed synthetic defect generation pipeline shown in Fig. 3, we leverage an LLM with visual captioning capabilities to create synthetic images with defects. The different tasks in this step are as follows:

*1) Template-based Caption Generation:* We chose GPT-4, which has been utilized for creating prominent visual-instruction datasets such as LLAVA, MIMIC IT, Objaverse, and Sceneverse [26]. Furthermore, from the literature, we found that LLM fine-tuned using rephrased samples achieved high accuracy. As demonstrated in Fig. 3, the process starts by taking a raw image as input and passing it through the LLM to generate a descriptive caption using a popular visual captioning technique known as the template-based caption generation approach. For our specific application, we developed a prompt template in collaboration with domain experts to capture essential visual defect characteristics, particularly those challenging to capture in real-life scenarios.

[3]Please see the animated version for better understanding.

*2) Rephrasing Algorithm for Diversity:* The template-based caption serves as input for rephrasing, as given in Algorithm 1. This technique transforms the prompt into new, intricate illustrations that accurately depict the specified elements missing in the original caption. By utilizing this procedure, we produce a significant number of synthetic examples, providing rich training examples for tuning the custom LLM. The Defect LLM dataset (DLLMDS) Generation Pipeline aims to solve the challenge of data scarcity in deploying LLM-based DT solutions.

**Lemma 1.** *We can define the problem as a constrained optimization problem where we want to maximize the diversity of the samples (D) and minimize the reconstruction loss (L).*

*Proof.* The objective function aims to maximize the diversity of the samples ($D$) and minimize the reconstruction loss ($L$). This can be represented as:

$$\underset{D,L}{\text{maximize}} \quad D - \lambda L \tag{1}$$

where:

- $D$ is the diversity of the samples,
- $L$ is the reconstruction loss,
- $\lambda$ is a trade-off parameter that balances the two objectives.

Each sample must be unique and more complex than the previous ones. The total number of samples must be less than or equal to $K$. These constraints can be represented as:

$$s_{\text{new}} \neq s_{\text{old}}, \ \forall s_{\text{old}} \in \text{Samples and } |\text{Samples}| \leq K \tag{2}$$

The Lagrangian for this problem is:

$$\mathcal{L}(D, L, \lambda, \mu) = D - \lambda L + \mu(K - |\text{Samples}|) \tag{3}$$

Where $\mu$ is the Lagrange multiplier for the constraint.

The optimality conditions for this problem are obtained by setting the partial derivatives of the Lagrangian with respect to $D$, $L$, and $\mu$ to zero:

$$\frac{\partial \mathcal{L}}{\partial D} = 1 - \mu = 0 \tag{4}$$

$$\frac{\partial \mathcal{L}}{\partial L} = -\lambda = 0 \tag{5}$$

$$\frac{\partial \mathcal{L}}{\partial \mu} = K - |\text{Samples}| = 0 \tag{6}$$

These conditions provide a mathematical proof for the algorithm's objective. The algorithm achieves this by generating unique and complex samples until it reaches the maximum number of samples ($K$), while ensuring that the reconstruction loss is minimized.

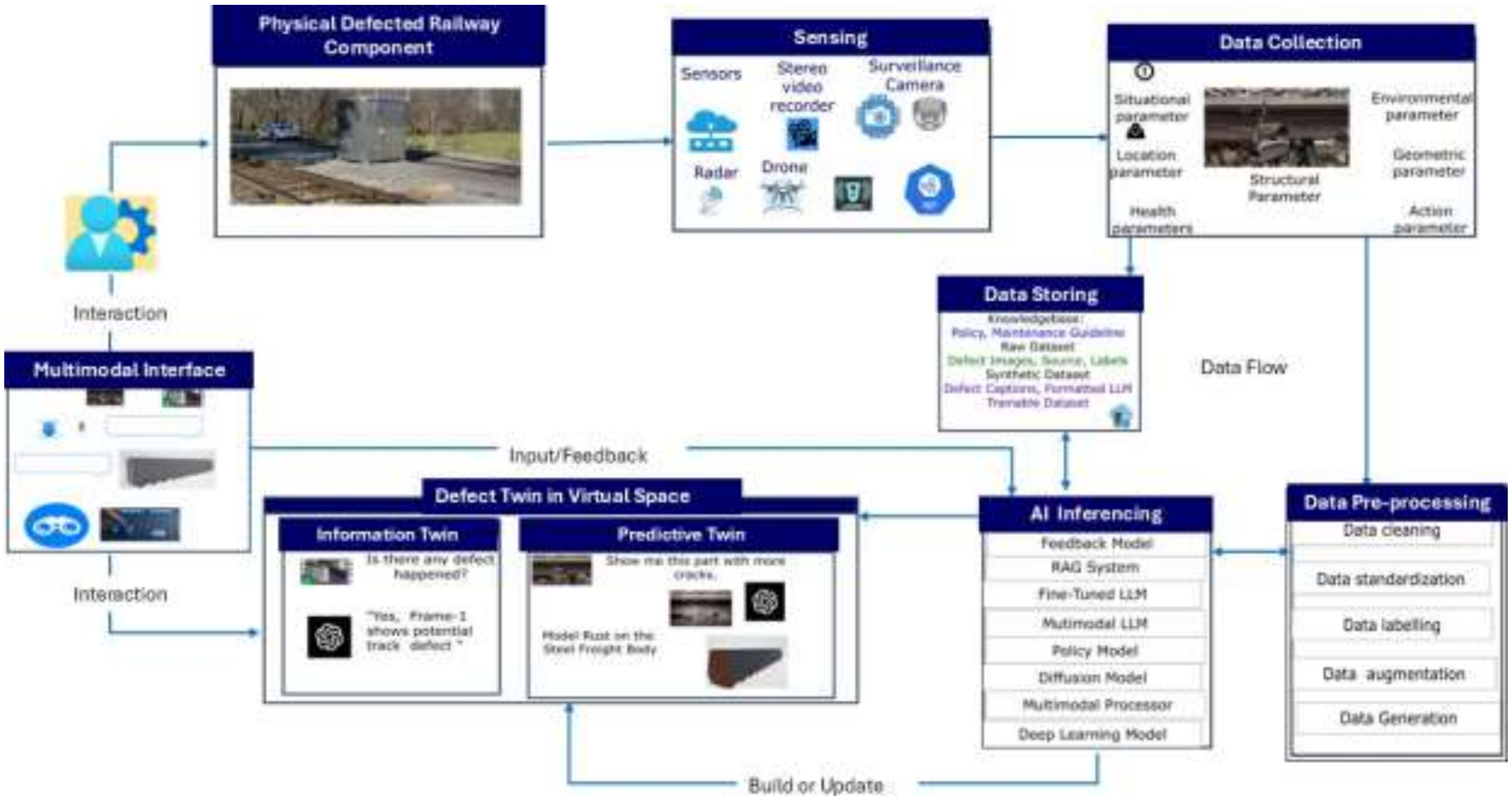

Fig. 1. High-level framework of LLM-based DefectTwin

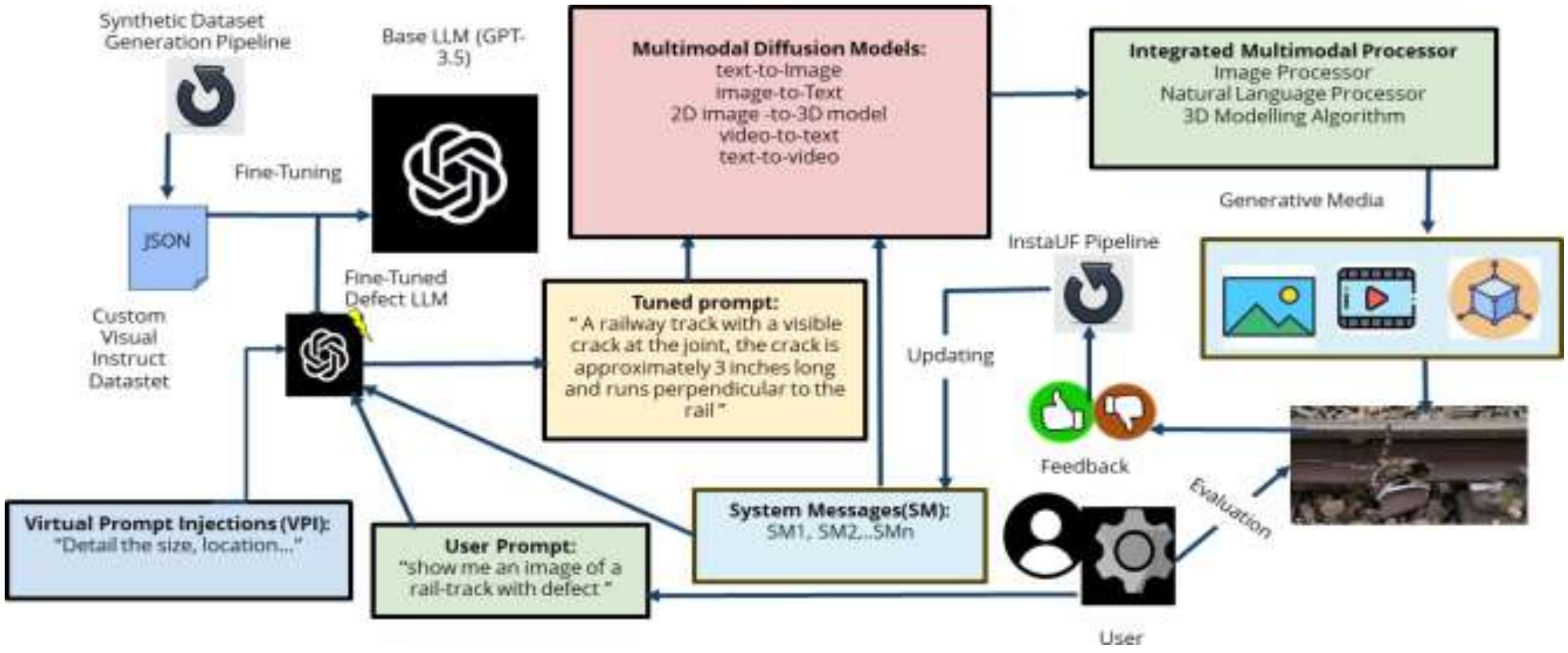

Fig. 2. The AI inferencing pipeline for railway defect detection.

□

The Algorithm 1, leverages template-based captions to generate a diverse and complex dataset. Each caption is expanded into multiple detailed samples using a language model, and each sample is paired with a system message to guide the fine-tuned DefectTwin LLM. Let us consider the following example:

Given a list of template-based captions: *captions = "A crack on the rail", "Corrosion at the joint", "A missing bolt"*

The algorithm generates multiple unique and complex samples for each caption. For example, for the caption "A crack on the rail", the generated samples could be:

*Samples = s1, s2, s3. where, s1 = "A crack 3 inches long on the rail surface, perpendicular to the track direction."; s2 = "A diagonal crack on the rail with a depth of 2mm, located near the joint."; s3 = "A longitudinal crack running along the rail track, extending 5 inches."*

Each sample is paired with a system message: *System Message = "Given the defect description provided, identify*

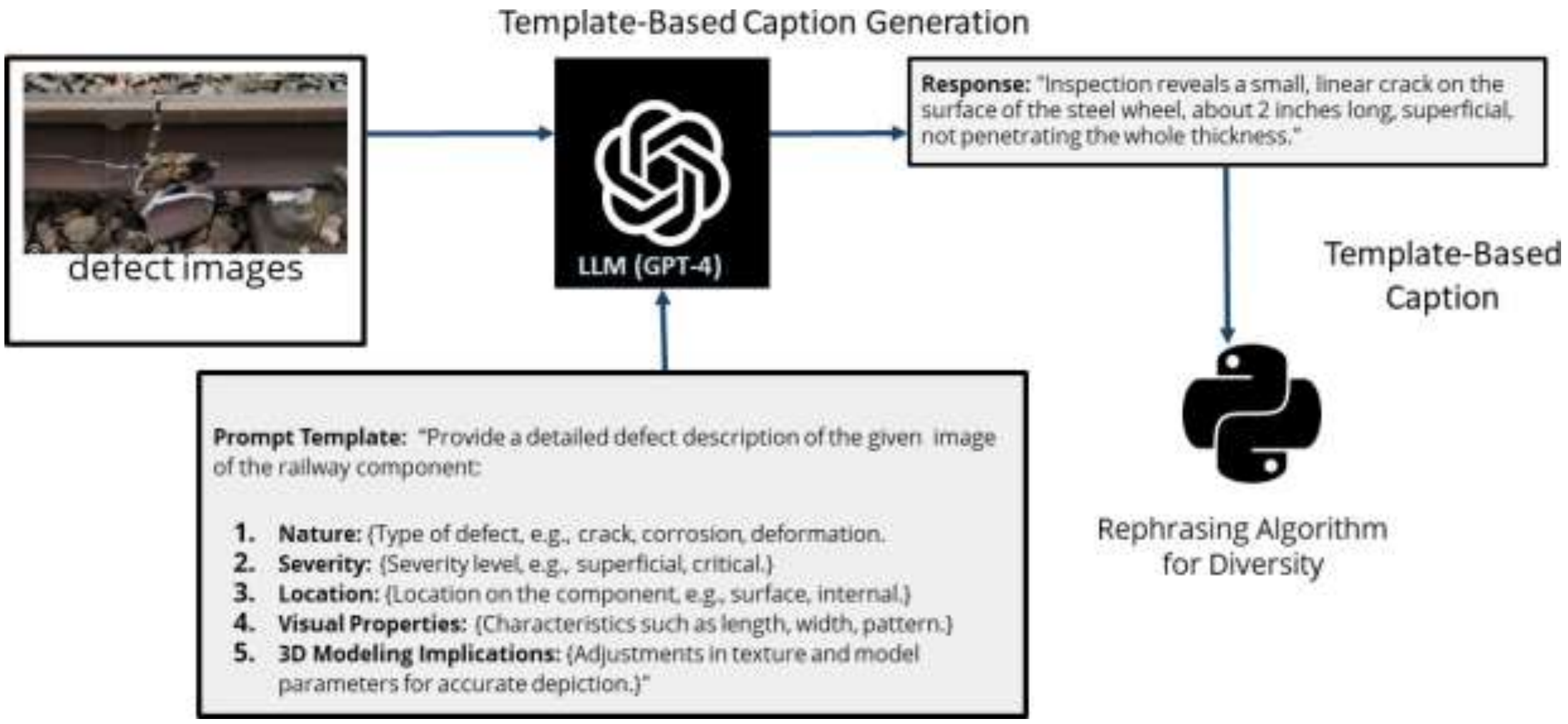

Fig. 3. Synthetic data generation pipeline.

**Algorithm 1** Rephrasing Algorithm for Diversity

**Input:** A list of template-based captions from defect images: `captions` = $[c_1, c_2, ..., c_n]$
**Output:** A dataset with multiple diverse and complex samples per caption, each accompanied by a system message.

**Step-1:** Create an empty list: `DS = []`
**Step-2:** For each caption ($c_i$) in `captions`, create an empty set for unique samples: `Samples = {}`
**Step-3:**
**while** the number of samples is less than $K$ **do**
  Generate a new sample ($s_{new}$) using a language model for $c_i$.
  Add $s_{new}$ to `Samples`.
**end while**
**Step-4:** For each generated sample ($s_{new}$) in `Samples`, formulate a system message.
Combine $s_{new}$ with the system message to form a structured data entry.
**Step-5:** Append all structured entries from `Samples` to `DS`. Ensure `DS` does not contain duplicates.

*potential risks and recommend preventive measures."* The diversity and complexity of samples are ensured by iterating until the number of unique samples, $S_{new}$, for each caption reaches a predefined limit $K$:

$$|S_{new}| < K \tag{7}$$

Each new sample $s_{new}$ is generated using a language model prompt: *"You are generating data to train an LLM. Based on the initial description: ci, create a prompt/response pair ensuring the response is more complex and diverse than previous ones."*

Unique and complex samples $s_{new}$ are added to the set Samples:

$$\text{if } (s_{new} \text{ is unique and complex}) \Rightarrow \text{add } s_{new} \text{ to Samples} \tag{8}$$

The dataset $DS$ is compiled by appending all structured entries from Samples, ensuring no duplicates in $DS$:

$$DS = DS \cup \text{Samples} \tag{9}$$

## B. Fine-Tuning the base LLM

The initial input to the fine-tuned LLM for DefectTwin consists of three elements: system messages (SM), user prompts, and Virtual Prompt injections (VPI). We describe each of these elements in detail.

As illustrated in Fig. 4, the process begins with a user providing a simple trigger scenario: "Steel wheel shows a radial crack." This scenario is processed by the fine-tuned LLM using system messages, user prompts, and the Visual Prompting Interface (VPI). System messages, such as "You are an expert railway component defect instructor," provide context. The user prompt describes the defect scenario. VPI adds details like location, size, and depth, e.g., "A radial crack, about two inches in length, is visible on the external circumference of the steel wheel".

The fine-tuned LLM integrates these inputs to create a comprehensive multimodal input. This is then passed to a Text-to-Image (TTI) model, generating a visual representation of the defect. The diffusion model produces an image showing the steel wheel with a visible radial crack, enhancing the realism and accuracy of the defect depiction (Fig. 4). The final output is more informative and precise, aiding in better visualization and understanding of the defect.

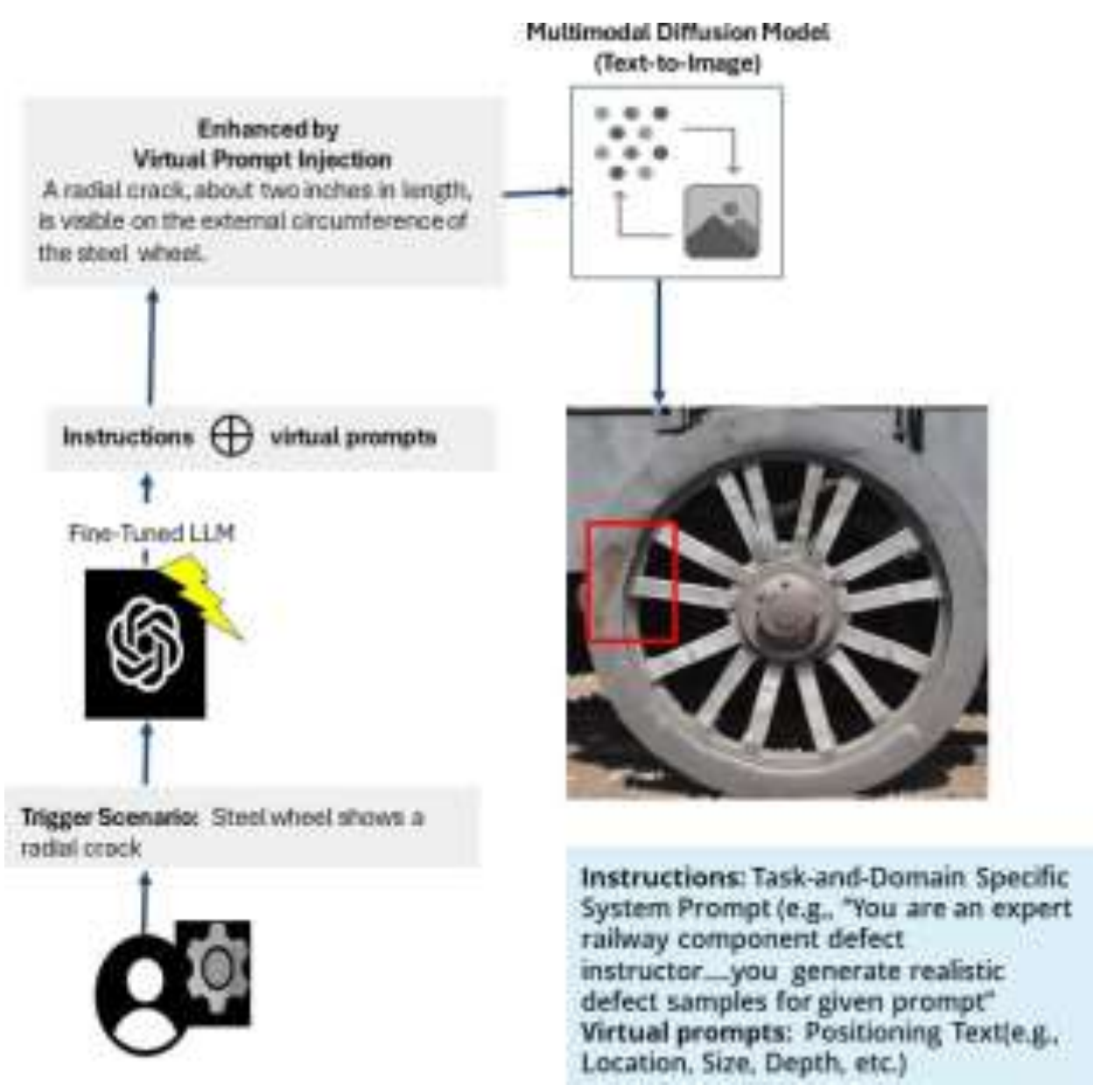

Fig. 4. The fine-tuned Defect LLM integrates system messages and VPI to generate a realistic depiction of a radial crack on a steel wheel.

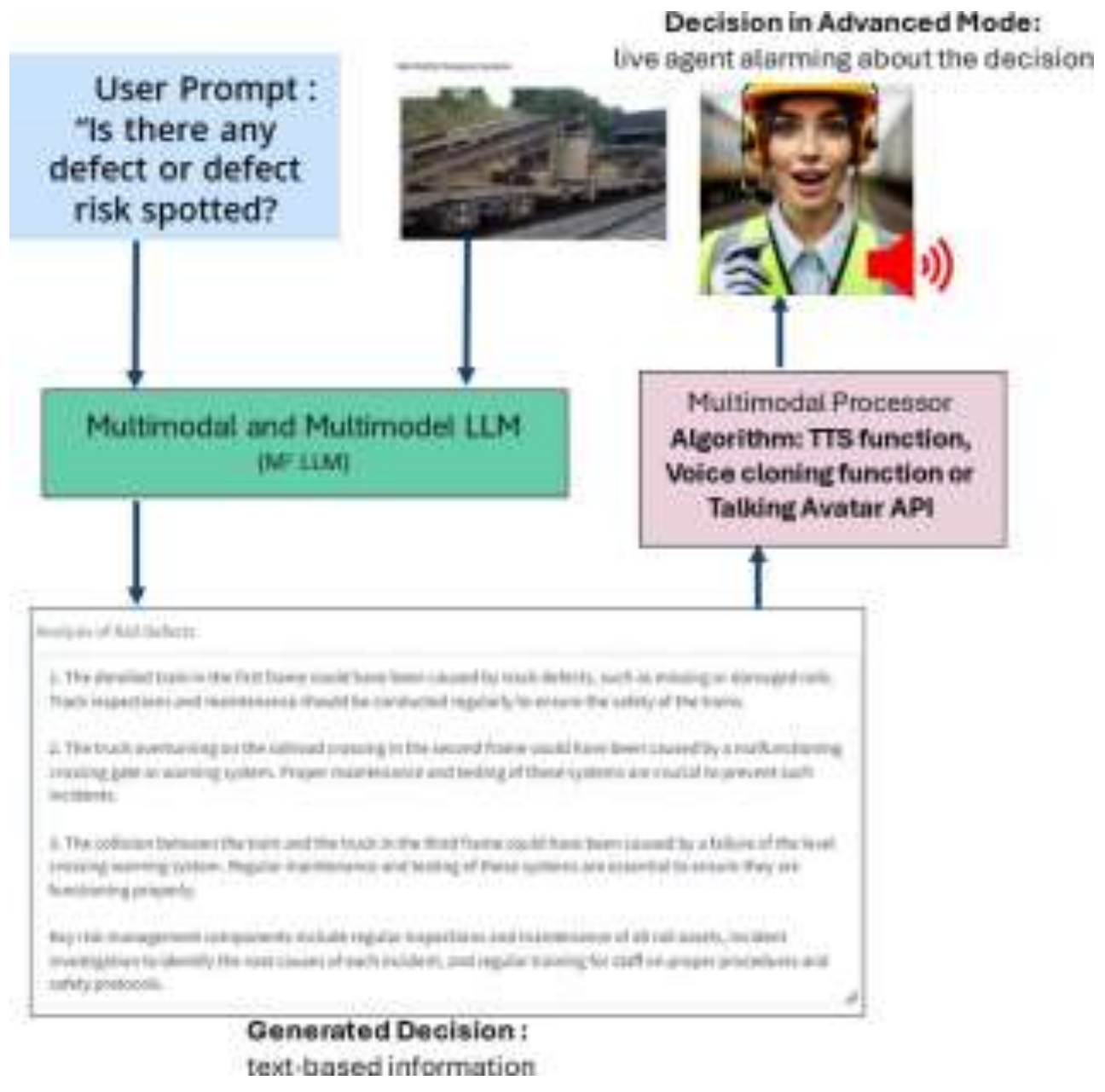

(a)

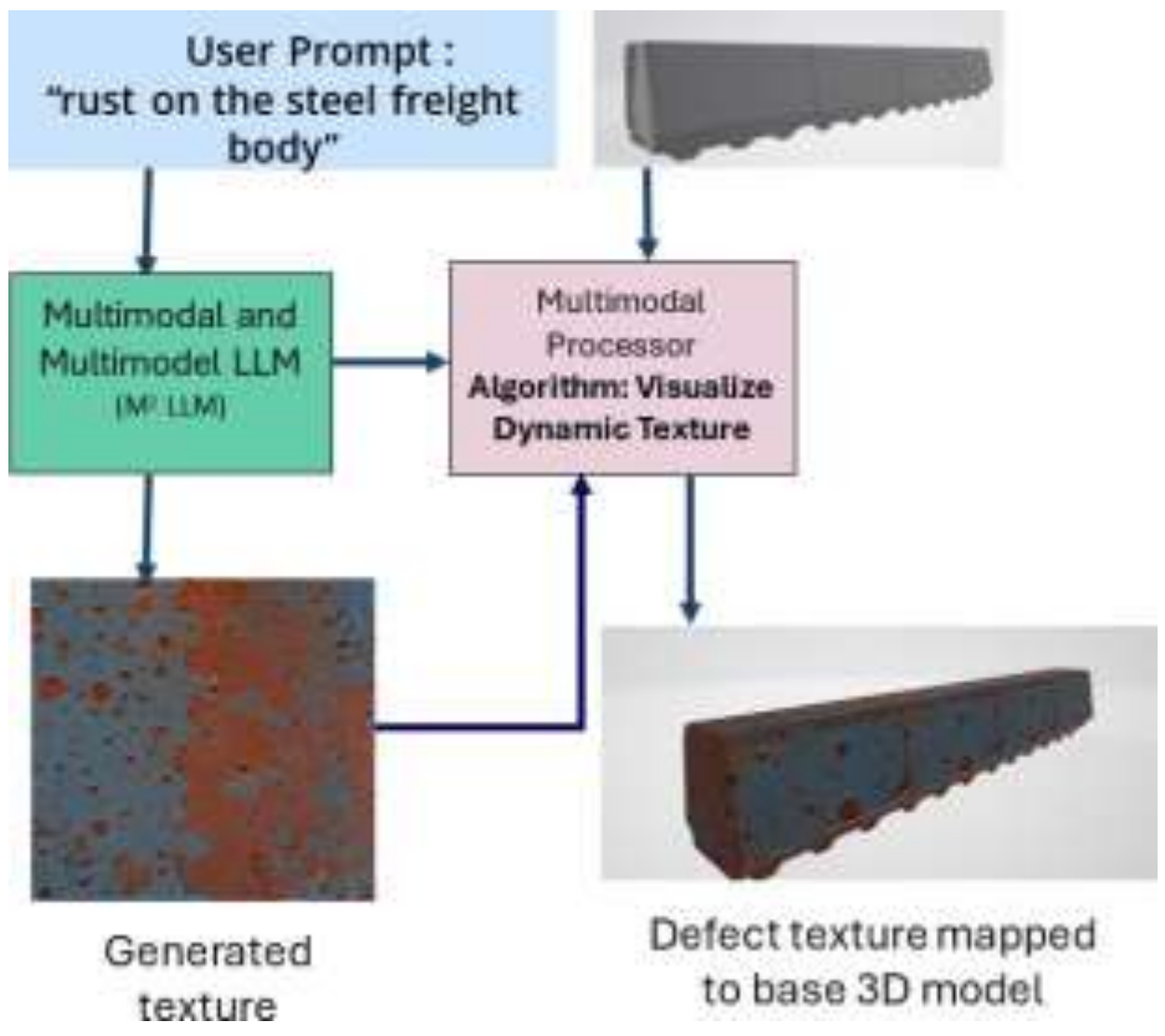

(b)

Fig. 5. Use of Multimodal processor in DefectTwin (a) Example I: Defect Analysis. (b) Example II: Predictive Visualization of Defect Characteristics.

## C. Multimodal Processing

The tuned prompt is passed to multimodal diffusion models, such as text-to-image, image-to-text, 2D image-to-3D, and video-to-text. These models generate images, videos, or 3D models that accurately depict the defects based on the tuned prompt. The output of these models is passed to the Multimodal Processor. The primary task of this unit is to take the output from the $M^2$ LLMs and transform it into a format that is consumable for the end-user. This involves interpreting various types of inputs, processing and transforming the generated data, and finally, outputting the results in a user-friendly manner. We discuss the general workflow of the Multimodal Processor concerning the two examples illustrated in Fig. 5 in the context of a DefectTwin system.

*1) Example 1 - Twining defect analysis process:* Let us consider an application to mimic the defect analysis of the railway defect. As illustrated in Fig. 5a, DefectTwin acts like an information twin by automating the decision-making process by analyzing the video stream. In this example, the AI inferencing engine receives multimodal inputs, including a video stream and a user prompt. Based on this input $M^2$ LLM in the DefectTwin framework make a decision. However, the decision might not be in a format that's easy for the user to understand. The Multimodal Processor converts this decision into a talking avatar, effectively communicating complex information in a user-friendly manner.

*2) Example 2- Texture Mapping and Visualization:* In this example, the $M^2$ LLM generates a defect texture based on the user prompt. However, this raw texture might not be directly usable. This is where the Multimodal Processor comes into play. As you can observe in Fig. 5b, a texture mapping algorithm is used to map the generated texture onto a base 3D model. This allows for dynamic visualization of defects in a simulated environment, enhancing the realism and usability of the data.

### D. AI User Interaction

Users interact with the fine-tuned LLM through a multi-modal interface and provide feedback on the system's performance. This feedback can vary as follows. 1) Positive or Negative Feedback. For example, positive feedback: "The LLM accurately identified the cracks in the railway track image." and negative feedback: "The LLM failed to identify the rust on the railway bolts." 2) Score-based Feedback. For example, 6 is scored on a scale of 1 to 10 for a response. Because it was able to identify major defects but missed out on minor ones. 3) Open-ended Feedback. E.g., Dissatisfaction: "You gave unrealistic defect". Refinement Need: "You should be able to differentiate between different types of defects such as cracks, rust, and mechanical wear." 4) Mixed Feedback - A score of 7 out of 10. "While it generally identifies major defects, it struggles with minor defects and often misses rust and small cracks". The user-interaction mechanism within DefectTwin is broken down as follows.

*1) Feedback Processing:* The feedback is input into an Instant user feedback (instaUF) handling pipeline that incorporates an instruct-tuned LLM designed to handle feedback. The instruct-tuned feedback LLM processes the feedback instantly. Let *( F )* represent the feedback, where *( F )*can be a score or textual feedback.

Based on the feedback, the system updates the message of the employed M² LLM. Let *( SM )* represent the system message output by the LLM, and $SM_{new}$) represent the updated message.

The update function can be represented as:

$$SM_{new} = Update(SM, F) \tag{10}$$

For example, if the LLM incorrectly identifies a crack in the railway track, the engineer might provide feedback as: a score of 1 out of 10, and a comment "Missed the small cracks". Based on this feedback the system message might be instructed to pay more attention to the size of the defect.

*2) Fine-Tuning Cycle:* The diversity of fine-tined LLM capabilities is required when the M² LLM cannot handle a specific type of defect, new samples are generated based on user feedback.

For example, if the LLM is fine-tuned on rust-based defects and incapable of handling mechanical defects like cracks or breaks, analyzing the user feedback new synthetic dataset is generated, and the current fine-tuned LLM is re-fine-tuned with new capability.

Each update builds on top of the previous model, retaining past improvements while incorporating new refinements. Let t represent the periodic update interval.

$$LLM_{t+1} = update(LLM_t, SM_{new}) \tag{11}$$

*3) instaUF for Optimization:* The instaUF pipeline is broken into two main components: the main function (see Algoritm 2) and the fine-tuning function (See Algorithm 3).

The main function handles the main loop of the algorithm, which collects user feedback, updates system parameters, and decides when to call the fine-tuning function. On the other hand, the fine-tuning function generates synthetic data using the DLLMDS pipeline and performs fine-tuning using the generated synthetic data to add more capabilities to the current fine-tuned LLM.

**Algorithm 2** Algorithm InstaUF - Main Function

**Input:**
- Fine-Tuned LLM ($LLM_i$), System Message (SM), Instruction (instruction)
- LLM settings parameters (LSP): Top-p ($p$), Top-k ($k$)
- Termination Criteria (tc), Fine-Tuning Interval (ft_interval)
- User Satisfaction Threshold (satisfaction_threshold)

**Output:**
- Fine-Tuned LLM ($LLM_{i+1}$) (only if fine-tuning occurs)
- Updated System Message (SM), Updated Instruction (instruction), Updated LSP

```
Initialize feedback vector (feedbacks) as an empty list.
Initialize iteration counter (counter) to 0.
Initialize user satisfaction score (satisfaction) to 100%.
while true do
    Collect user feedback (F).
    Append F to feedback vector (feedbacks).
    Process feedback using the Feedback Processing Function: (SM, instruction, p', k', satisfaction) = Update(SM, F, instruction).
    Update system parameters (SM, instruction, p', k').
    if satisfaction ¡ satisfaction_threshold or counter reaches ft_interval then
        Call Fine-Tune Function (LLM, D_s, p, k, feedbacks)
        Reset feedback vector (feedbacks).
        Reset satisfaction to 100% if it was below threshold.
        Reset counter to 0 if interval was reached.
    end if
    Increment counter by 1.
    if tc is met then
        break
    end if
end while
Return Updated LLM (only if fine-tuning occurred), SM, instruction, and LSP (p', k')
```

**Algorithm 3** Algorithm InstaUF - Fine-Tuning Function

**Input:**
- LLM, Synthetic Dataset ($D_s$), Top-p ($p$), Top-k ($k$), Feedback Vector (feedbacks)

**Output:**
- Fine-Tuned LLM ($LLM_{i+1}$)

```
Generate synthetic dataset (D_s) using DLLMDS pipeline.
Fine-tune the LLM using (D_s): FineTune(LLM, D_s, p, k).
Set LLM_{i+1} to the fine-tuned LLM.
Return (LLM_{i+1})
```

**Lemma 2.** *The Greedy Algorithm for Fine-Tuning LLM with user feedback maximizes user satisfaction (S) at or near 100% and minimizes the number of fine-tuning operations (FT).*

*Proof.* We want to maximize $S$ subject to the constraint that $FT$ is minimized. We can formulate this as a constrained optimization problem:

$$\begin{aligned} \underset{S,FT}{\text{maximize}} \quad & S \\ \text{subject to} \quad & FT \leq T, \end{aligned} \tag{12}$$

where $T$ is the total number of iterations.

The Lagrangian for this problem is:

$$L(S, FT, \lambda) = S - \lambda(FT - T), \tag{13}$$

where $\lambda$ is the Lagrange multiplier.

Taking the partial derivatives and setting them equal to zero gives the following conditions:

$$\frac{\partial L}{\partial S} = 1 - 0 = 1, \tag{14}$$

$$\frac{\partial L}{\partial FT} = -\lambda = 0, \tag{15}$$

$$\frac{\partial L}{\partial \lambda} = FT - T = 0. \tag{16}$$

**Main Function:** The main function can be represented by the following iterative equation:

$$S^t = \begin{cases} 100 & \text{if } t\%\alpha = 0 \text{ or } S_{t-1} < \beta \\ f(S_{t-1}, F_t) & \text{otherwise} \end{cases} \tag{17}$$

where:

- $S_t$ is the user satisfaction score at iteration $t$,
- $F_t$ is the feedback at iteration $t$,
- $f(S_{t-1}, F_t)$ is the feedback processing function,
- $\alpha$ = ft_interval is the fine-tuning interval,
- $\beta$ = satisfaction_threshold is the satisfaction threshold.

**Fine-Tuning Function:** The fine-tuning function can be represented by the following equation:

$$LLM_{i+1} = \text{FineTune}(LLM_i, D_s, p, k) \tag{18}$$

where:

- $LLM_i$ is the fine-tuned LLM at iteration $i$,
- $D_s$ is the synthetic dataset,
- $p$ and $k$ are the top-p and top-k parameters,
- FineTune$(LLM_i, D_s, p, k)$ is the fine-tuning function.

From equation (15), $\lambda = 0$ implies $FT \leq T$ is not binding, so the algorithm does not always fine-tune every iteration. While, from equation (16), $FT = T$ implies the algorithm fine-tunes exactly $T$ times, contradicting the earlier finding. The algorithm balances maximizing user satisfaction and minimizing fine-tuning, resetting satisfaction to 100% after fine-tuning and maintaining it above a threshold. It limits fine-tuning to necessary times and regular intervals, forming the core logic of Algorithms 3 and 4.

□

## IV. Experiment and Analysis

### A. Data

In this research, we have employed both original [4] and synthetic data[5] to evaluate the performance and usefulness of DefectTwin.

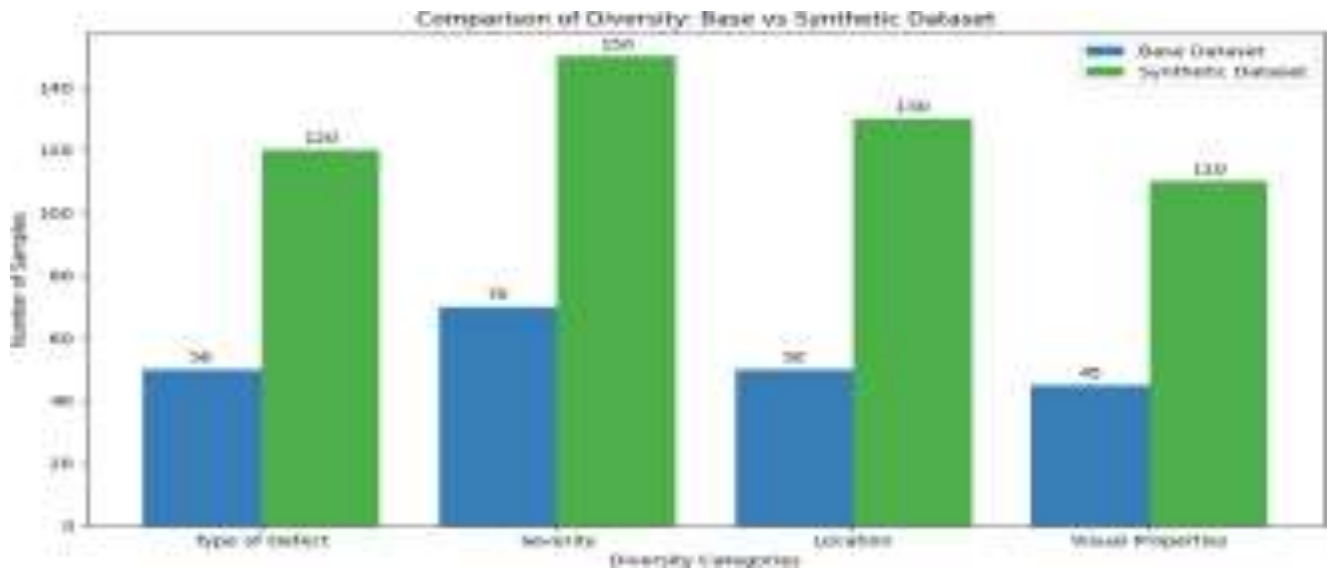

Fig. 6. Sample diversity achieved by employing DLLMDS pipeline.

- **Raw Data:** We utilized two primary datasets: the Canadian Pacific Railway (CPR) Defect Dataset with 1000 defect images and 26 unique labels, and an Expanded General Category Dataset with 150 samples of various damaged railway components collected from open-source images. The latter addresses the CPR dataset's limited diversity and preprocessed format.
- **Synthetic Datasets for Fine-Tuning:** To enhance our datasets, we used the DLLMDS pipeline to generate the Defect visual-instruct dataset containing visual instructions and responses related to defects and maintenance, and the Texture visual-instruct dataset, providing defect texture visual response data.
- **Test Data for Accuracy and Response Generation Evaluation:** We evaluated our defect twin framework using image-based defect detection with 100 images each from the CPR and expanded datasets, video-based defect identification with data from CPR and open-source datasets, and response generation performance with text prompts and multimodal inputs from publicly available YouTube videos.

### B. Evaluation Parameters

We utilized various evaluation metrics to assess the $M^2$ LLM-based AI inferencing components of DefectTwin. For the defect detection task, we measured accuracy using Precision, Recall, F1-score, and AUC (Area Under the Curve). To evaluate the relevance of the generated responses, we employed Answer Relevance, Context Relevance, and ROUGE-L Score. For assessing optimality, we measured latency and the number of tokens generated. To determine the usefulness, we used a scale of 1 to 10, following the approach presented in [9].

### C. Experiment Setup

We developed the DefectTwin apps on Huggingface. The Gradio framework facilitated an interactive interface for seam-

[4]Please download the kaggle dataset from here

[5]Please download our open-source dataset from here

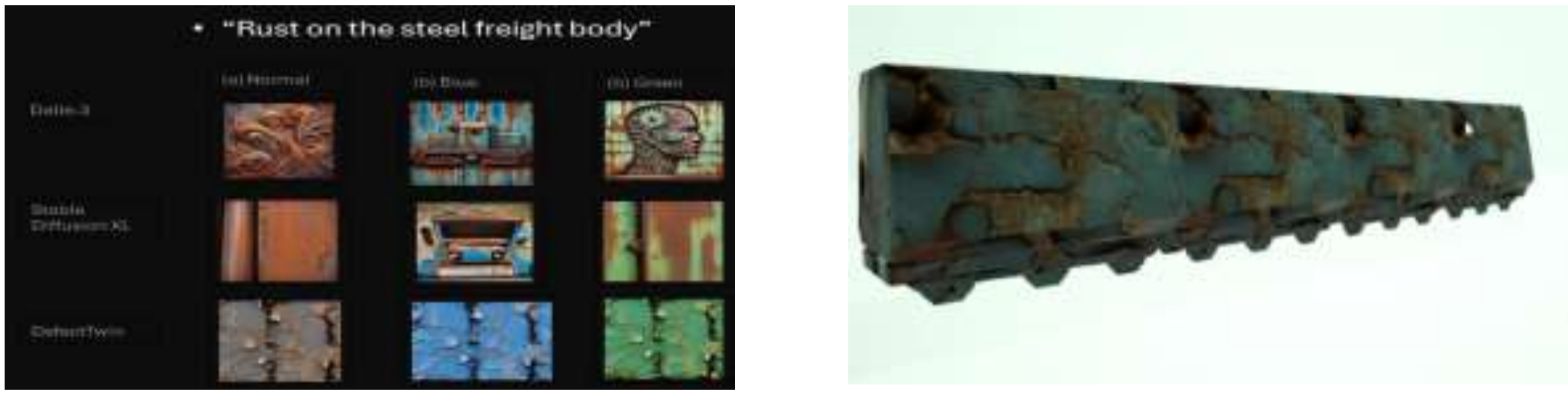

(a) (b)

Fig. 7. Comparative Analysis of Rust Texture Simulation on Steel Freight Bodies Across Different Generative Models. (a)Model Comparisons for Simulating Rust Textures in Varied Colors (b)3D Visualization of Rust Impact on Steel Freight Body

TABLE I
PERFORMANCE METRICS OF MODELS ON DIFFERENT MEDIA

*(a) Performance on Image*

| Model | In-Domain (Track, Assets) | | | | Zero-Shot (Infrastructure) | | | |
|---|---|---|---|---|---|---|---|---|
| | Precision | Recall | F1-score | AUC | Precision | Recall | F1-score | AUC |
| Instruct-BLIP | 0.55 | 0.58 | 0.57 | 0.58 | 0.35 | 0.4 | 0.37 | 0.38 |
| LLAVA-Instruct | 0.85 | 0.86 | 0.85 | 0.86 | 0.45 | 0.45 | 0.47 | 0.48 |
| GPT-4o | 0.68 | 0.64 | 0.62 | 0.83 | 0.4 | 0.45 | 0.42 | 0.43 |
| Gemini-Pro-Vision | 0.88 | 0.88 | 0.88 | 0.89 | 0.48 | 0.5 | 0.49 | 0.5 |
| Proposed Model | 0.92 | 0.93 | 0.92 | 0.93 | 0.6 | 0.65 | 0.62 | 0.63 |

*(b) Performance on Video*

| Model | In-domainn(Track, Assets) | | | | Zero-Shot | | | |
|---|---|---|---|---|---|---|---|---|
| | Precision | Recall | F1-score | AUC | Precision | Recall | F1-score | AUC |
| Instruct-BLIP | 0.3 | 0.35 | 0.32 | 0.34 | 0.2 | 0.25 | 0.22 | 0.24 |
| GPT-4o | 0.65 | 0.68 | 0.65 | 0.67 | 0.52 | 0.52 | 0.51 | 0.53 |
| LLAVA | 0.35 | 0.4 | 0.37 | 0.39 | 0.25 | 0.3 | 0.27 | 0.29 |
| Gemini-Pro-Vision | 0.71 | 0.72 | 0.71 | 0.73 | 0.45 | 0.48 | 0.45 | 0.47 |
| Proposed Model | 0.76 | 0.74 | 0.77 | 0.77 | 0.55 | 0.58 | 0.55 | 0.57 |

less input of textual and visual data, providing a comprehensive platform to evaluate the performance of $M^2$ LMM in handling multimodal tasks. The apps were tested on an ipad-10th generation with a capacity of 64 GB memory, 10.9 inch multi-touch display with ips technology, A14 bionic chip, 6-core CPU, 4-core graphics, and 16-core neural engine. In our experiments, we evaluated both unimodal (GPT-3.5) [28] and multimodal LLMs (Instruct-BLIP [29], GPT-4 [10], LLAVA [24], Gemini-Pro-Vision [11], and our Proposed Model). The unimodal LLM handled text-based tasks, while the multimodal models processed images and videos alongside text, enhancing accuracy and contextual relevance.

We selected models based on their ability to manage defect descriptions, case-based scenarios, and maintenance contexts, considering both in-domain and zero-shot generalizability. For zero-shot evaluation, railway infrastructures like bridges and stations served as out-of-domain components, while defect categories included wheels, gates, doors, rail surfaces, and tracks.

### D. Ablation Study

We conducted several ablations to evaluate the impact of DefectTwin in the context of CE and railway defect inspection incorporating multimodal data.

*1) Diversity (Does DefectTwin Achieve Diversity in Synthetic Examples?):* The proposed synthetic dataset generation approach for fine-tuning significantly enhances the diversity in defect-specific characteristics compared to the basic visual captioning method. As illustrated in Fig. 6, the DLLMDS pipeline effectively captures a wide range of defect characteristics in the synthetic dataset, surpassing the coverage of the base dataset generated through simple visual captioning. The impact of fine-tuning on texture generation for defect visualization is shown in the ablation study where ’rust on steel’ was generated in three colors—Normal, Blue, and Green (Fig. 7).

Fig. 7a displays outputs from three models: Dalle-3, Stable DiffusionXL, and DefectTwin. While Dalle-3 and Stable DiffusionXL produced stylized rust effects, DefectTwin generated realistic rust textures with intricate details and variegated coloration, closely resembling real-world rust patterns. The fine-tuning process captured the defect textures visibly (Fig. 7b), providing high-fidelity visual simulations beneficial for maintenance planning and predictive diagnostics.

*2) Generalizability (Does DefectTwin identify unseen data and classes with high precision and consistency?):* Our proposed model, DefectTwin, demonstrates robust accuracy across multimodal data, excelling in both image and video-based defect detection. For image data, DefectTwin achieved

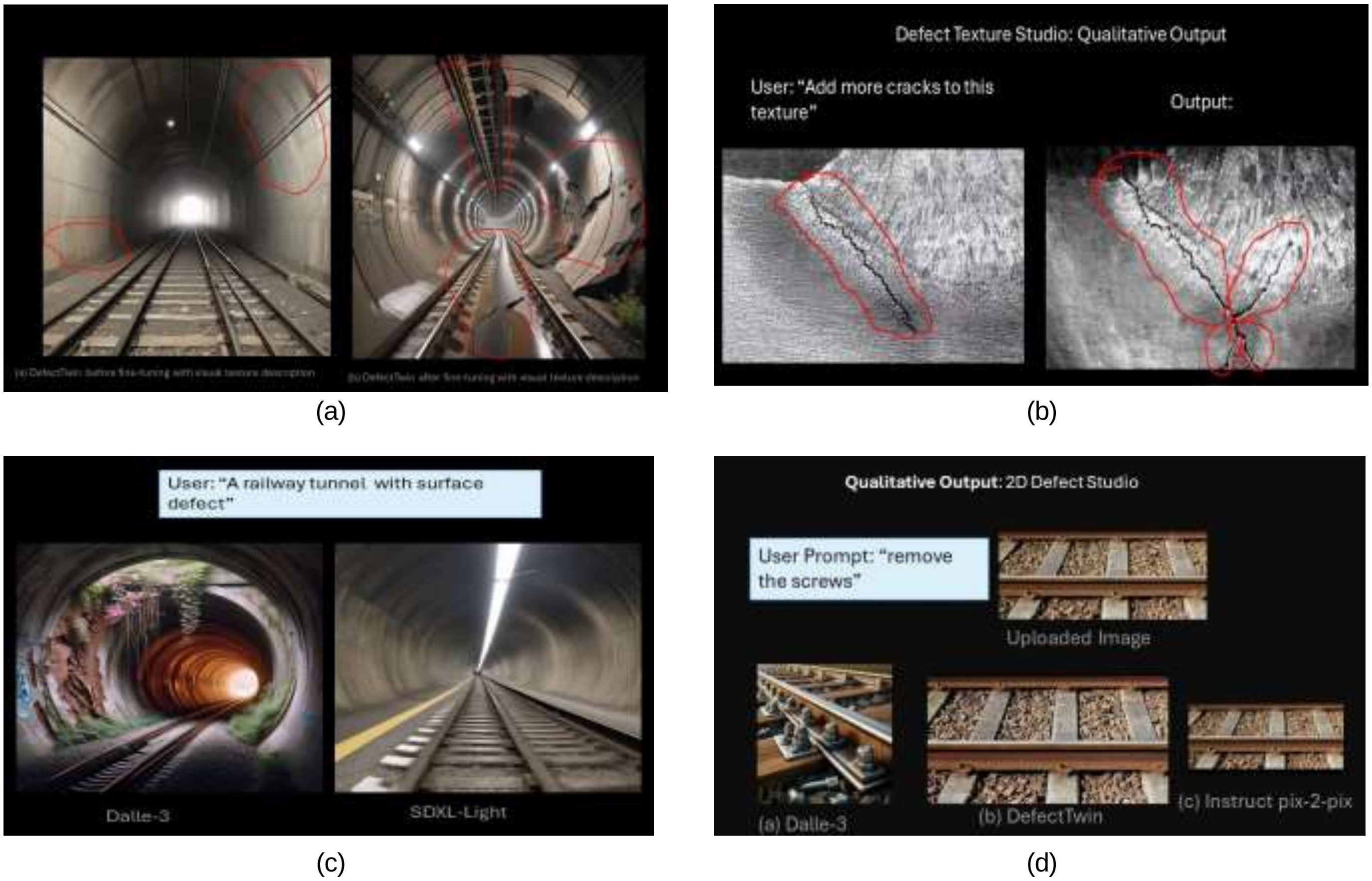

Fig. 8. (a) Surface Defect Visualization by DefectTwin (b) Surface Defect Visualization by Existing Model (c) Screw Removal Prompt (d) Crack Enhancement Prompt

top scores in in-domain scenarios with a precision of 0.92, recall of 0.93, F1-score of 0.92, and AUC of 0.93 (Table Ia). It also showed adaptability in zero-shot scenarios with a precision of 0.6 and an F1 score of 0.62. In video data, DefectTwin maintained high performance in familiar contexts with a precision of 0.76, recall of 0.74, and F1-score of 0.77, though performance declined in zero-shot scenarios to a precision and F1-score of 0.55, highlighting the challenges of video analysis (Table Ib). This comprehensive evaluation confirms DefectTwin's strong accuracy and capability in handling both familiar and novel environments across multimodal data.

*3) QoE (Does DefectTwin Adapt Based on User Needs):* We conducted this ablation based on the following examples:

- **Surface Defect Visualization:** Initially, DefectTwin's visualization of "a railway tunnel with surface defect" lacked detail. After fine-tuning, it produced detailed and realistic textures accurately depicting surface defects (Fig.8a). In contrast, Dalle-3 and SDXL-Light provided visually appealing but less accurate results (Fig. 8b).
- **Crack Enhancement Prompt:** Before fine-tuning, DefectTwin's output for adding cracks was not detailed enough (Fig.8c). Post-fine-tuning, it accurately added pronounced and detailed crack textures, significantly enhancing defect visibility. Compared to this, other models struggled to achieve the same level of detail.

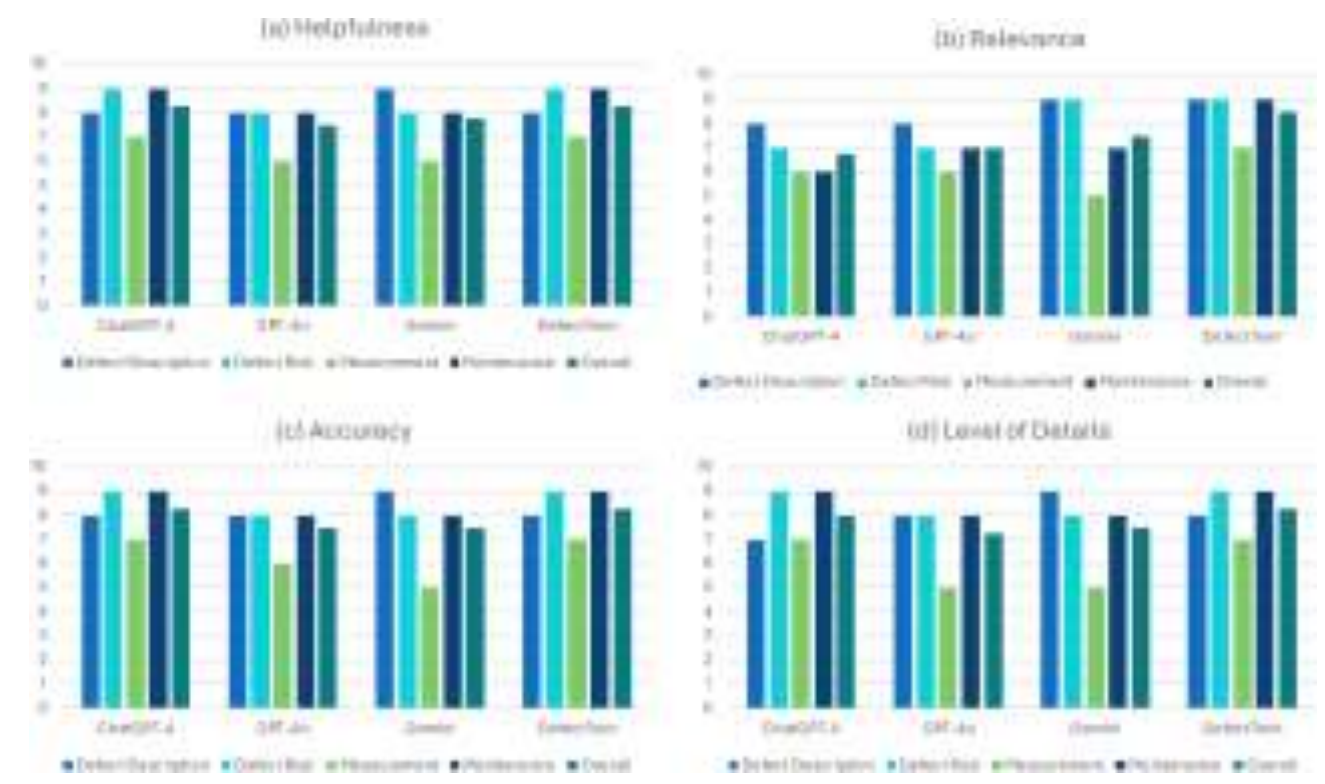

Fig. 9. Response utility for rail defect inspection.

- **Screw Removal Prompt:** DefectTwin effectively removed screws while maintaining the integrity of the underlying image (Fig.8d), demonstrating superior performance. Dalle-3 provided an aesthetically pleasing but less accurate response, while Instruct pix-2-pix offered more practical outputs but was not as precise as DefectTwin.

*4) Usefuleness and Relevance (Does DefectTwin generate useful and relevant responses?):*

- **Usefulness** The Fig. 9 illustrates that DefectTwin out-

TABLE II
ANSWER RELEVANCE ACROSS DIFFERENT TASK: MULTIMODAL

*(a) Defect Detection*

| Model | Answer Relevance | Context Relevance | ROUGE-L Score |
|---|---|---|---|
| Instruct-BLIP | 0.65 | 0.37 | 0.31 |
| GPT-4o | 0.43 | 0.52 | 0.54 |
| LLAVA | 0.45 | 0.58 | 0.36 |
| Gemini-Pro-vision | 0.41 | 0.51 | 0.21 |
| Proposed | 0.79 | 0.97 | 0.94 |

*(b) Defect Risk Identification*

| Model | Answer Relevance | Context Relevance | ROUGE-L Score |
|---|---|---|---|
| Instruct-BLIP | 0.17 | 0.20 | 0.22 |
| GPT-4o | 0.36 | 0.49 | 0.32 |
| LLAVA | 0.59 | 0.52 | 0.21 |
| Gemini-Pro-Vision | 0.31 | 0.67 | 0.25 |
| Proposed | 0.78 | 0.95 | 0.92 |

*(c) Maintenance Recommendation*

| Model | Answer Relevance | Context Relevance | ROUGE-L Score |
|---|---|---|---|
| Instruct-BLIP | 0.35 | 0.22 | 0.476 |
| GPT-4o | 0.51 | 0.35 | 0.62 |
| LLAVA | 0.27 | 0.43 | 0.33 |
| Gemini-Pro-Vision | 0.35 | 0.53 | 0.28 |
| Proposed Model | 0.46 | 0.86 | 0.84 |

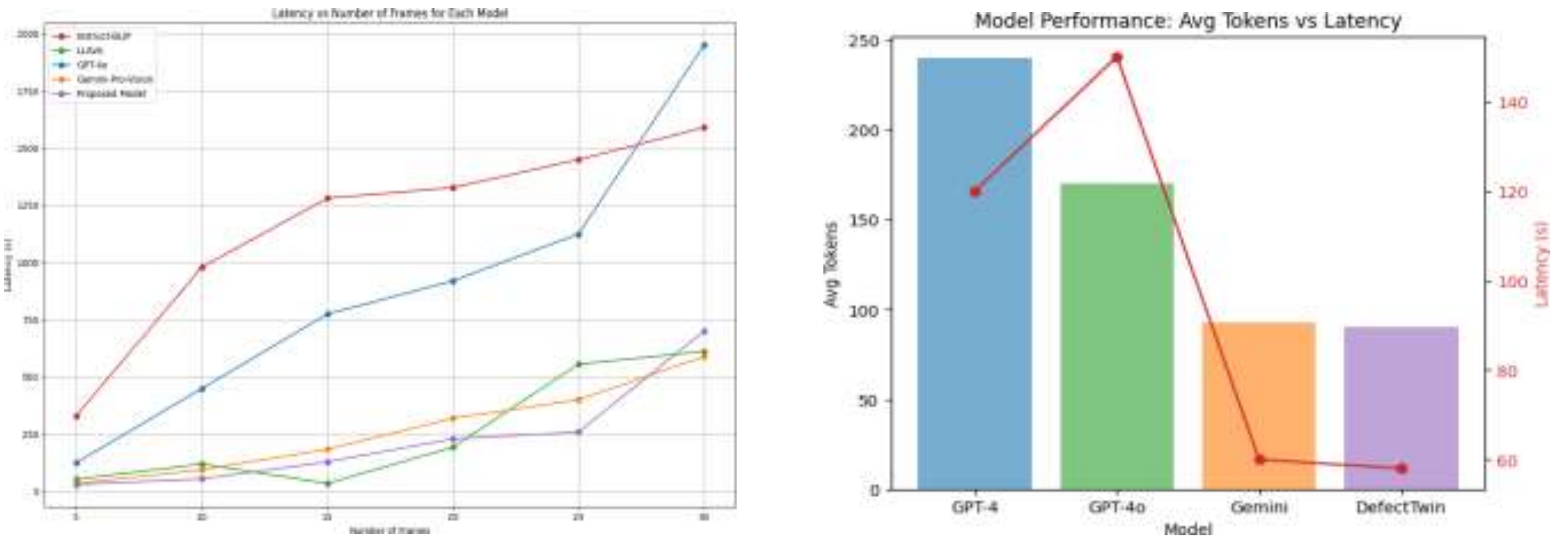

(a) (b)

Fig. 10. (a) Latency and Token Generation: Text-To-Text (b) Latency VS Number of Video Frames Processing

performed existing models in terms of helpfulness, relevance, accuracy, and level of detail scoring above 8 out of most of the criteria.

- **Relevance** Table II demonstrates that comparatively DefectTwin performed better than related models in most of the tasks. However, relevance decreased significantly for measurement or maintenance-related tasks. Because the fine fine-tuning focused more on defect description rather than monitoring.

*5) Optimality (Does DefectTwin offer optimality to be included in CE products?):* Performance on Token Generation and Latency: DefectTwin achieves an optimal balance with a moderate number of tokens and lower latency compared to GPT-4O (see Fig. 10b), maintaining processing efficiency essential for real-time applications in consumer electronics.

DefectTwin exhibits a consistent but lower increase in latency with more frames than Instruct-BLIP [18] and LLAVA (see Fig. 10a), demonstrating better scalability for continuous data streams like video monitoring in consumer electronics.DefectTwin's efficient token management reduces computational load and minimizes irrelevant responses, crucial for accurate and fast processing in consumer electronics systems with limited resources.

## V. CONCLUSION AND FUTURE WORK

In this paper, we introduced DefectTwin, integrating LLMs with DT technology to enhance visual defect inspection in railway components. Our multimodal and M² AI inferencing pipeline achieved high precision in defect identification. DefectTwin faces challenges in handling maintenance-related prompts. Therefore, in further extension of this work, we aim to improve synthetic data generation and explore RAG-based approaches for better accuracy and reliability. Although we obtained satisfactory usefulness through LLM-based evaluation, detailed user evaluation with domain-specific stakeholders still needs to be addressed. Based on the outcomes of this research, we anticipate DefectTwin as a potential solution to be extended in similar fields like- automotive and aerospace.

## References

[1] S. Sai, A. Rastogi, and V. Chamola, "Digital twins for consumer electronics," *IEEE Consumer Electronics Magazine*, 2023.

[2] J. Liu and C. Zhang, "Effective analysis and intelligent decision making of consumer electronics data based on machine learning under smart city," *IEEE Transactions on Consumer Electronics*, 2023.

[3] S. Ghaboura, R. Ferdousi, F. Laamarti, C. Yang, and A. El Saddik, "Digital twin for railway: A comprehensive survey," *IEEE Access*, vol. 11, pp. 120237–120257, 2023.

[4] R. Ferdousi, F. Laamarti, C. Yang, and A. E. Saddik, "A reusable ai-enabled defect detection system for railway using ensembled cnn," *arXiv preprint arXiv:2311.14824*, 2023.

[5] V. Caˆmara, R. Mendonca-Neto, A. Silva, and L. Cordovil, "A large language model approach to sql-to-text generation," in *2024 IEEE International Conference on Consumer Electronics (ICCE)*, pp. 1–4, 2024.

[6] K. Suzuki, J. Cai, J. Li, T. Yamauchi, and K. Tei, "A comparative evaluation on melody generation of large language models," in *2023 IEEE International Conference on Consumer Electronics-Asia (ICCE-Asia)*, pp. 1–4, 2023.

[7] H. Chung, S. Hyun, and Y.-G. Ha, "Battlefield situation awareness using pretrained generative llm," in *2024 IEEE International Conference on Big Data and Smart Computing (BigComp)*, pp. 397–398, 2024.

[8] J. Yan, V. Yadav, S. Li, L. Chen, Z. Tang, H. Wang, V. Srinivasan, X. Ren, and H. Jin, "Backdooring instruction-tuned large language models with virtual prompt injection," in *Proceedings of the 2024 Conference of the North American Chapter of the Association for Computational Linguistics: Human Language Technologies (Volume 1: Long Papers)*, pp. 6065–6086, 2024.

[9] H. Liu, C. Li, Q. Wu, and Y. J. Lee, "Visual instruction tuning," *Advances in neural information processing systems*, vol. 36, 2024.

[10] P. Taveekitworachai, M. C. Gursesli, F. Abdullah, S. Chen, F. Cala, A. Guazzini, A. Lanata, and R. Thawonmas, "Journey of chatgpt from prompts to stories in games: the positive, the negative, and the neutral," in *2023 IEEE 13th International Conference on Consumer Electronics - Berlin (ICCE-Berlin)*, pp. 202–203, 2023.

[11] G. Team, R. Anil, S. Borgeaud, Y. Wu, J.-B. Alayrac, J. Yu, R. Soricut, J. Schalkwyk, A. M. Dai, A. Hauth, *et al.*, "Gemini: a family of highly capable multimodal models," *arXiv preprint arXiv:2312.11805*, 2023.

[12] H. Azzuni, S. Jamal, and A. Elsaddik, "utalk: Bridging the gap between humans and ai," in *2024 IEEE International Conference on Consumer Electronics (ICCE)*, pp. 1–4, 2024.

[15] C. K. Wu, C.-T. Cheng, Y. Uwate, G. Chen, S. Mumtaz, and K. F. Tsang, "State-of-the-art and research opportunities for next-generation consumer electronics," *IEEE Transactions on Consumer Electronics*, vol. 69, no. 4, pp. 937–948, 2022.

[13] R. Ferdousi, C. Yang, M. A. Hossain, F. Laamarti, M. S. Hossain, and A. E. Saddik, "Generative model-driven synthetic training image generation: An approach to cognition in railway defect detection," *Cognitive Computation*, pp. 1–16, 2024.

[14] A. El Saddik, F. Laamarti, and M. Alja'Afreh, "The potential of digital twins," *IEEE Instrumentation & Measurement Magazine*, vol. 24, no. 3, pp. 36–41, 2021.

[16] W. Liu, X. Xu, L. Qi, X. Zhou, H. Yan, X. Xia, and W. Dou, "Digital twin-assisted edge service caching for consumer electronics manufacturing," *IEEE Transactions on Consumer Electronics*, 2024.

[17] P. Bhattacharya, V. K. Prasad, A. Verma, D. Gupta, A. Sapsomboon, W. Viriyasitavat, and G. Dhiman, "Demystifying chatgpt: An in-depth survey of openai's robust large language models," *Archives of Computational Methods in Engineering*, pp. 1–44, 2024.

[18] T. Brooks, A. Holynski, and A. A. Efros, "Instructpix2pix: Learning to follow image editing instructions," in *Proceedings of the IEEE/CVF Conference on Computer Vision and Pattern Recognition*, pp. 18392–18402, 2023.

[19] J. Zhang, Y. Zhang, M. Chu, S. Yang, and T. Zu, "A llm-based simulation scenario aided generation method," in *2023 IEEE 7th Information Technology and Mechatronics Engineering Conference (ITOEC)*, vol. 7, pp. 1350–1354, 2023.

[20] J. Jeon, B. Jeong, and Y.-S. Jeong, "Intelligent resource scaling for container based digital twin simulation of consumer electronics," *IEEE Transactions on Consumer Electronics*, 2023.

[21] R. Rassin, S. Ravfogel, and Y. Goldberg, "Dalle-2 is seeing double: flaws in word-to-concept mapping in text2image models," *arXiv preprint arXiv:2210.10606*, 2022.

[22] W. Ma, C. Yang, and C. Ka¨stner, "(why) is my prompt getting worse? rethinking regression testing for evolving llm apis," in *2024 IEEE/ACM 3rd International Conference on AI Engineering – Software Engineering for AI (CAIN)*, pp. 166–171, 2024.

[23] S. Parida, S. Sekhar, S. Panda, S. Jena, A. Parida, S. K. Sahoo, and S. R. Dash, "Olive: An instruction following llama model for odia language," in *2023 IEEE Silchar Subsection Conference (SILCON)*, pp. 1–7, 2023.

[24] J. Huang, J. Zhang, K. Jiang, H. Qiu, and S. Lu, "Visual instruction tuning towards general-purpose multimodal model: A survey," *arXiv*, 2023.

[25] M. Deitke, D. Schwenk, J. Salvador, L. Weihs, O. Michel, E. VanderBilt, L. Schmidt, K. Ehsani, A. Kembhavi, and A. Farhadi, "Objaverse: A universe of annotated 3d objects," in *Proceedings of the IEEE/CVF Conference on Computer Vision and Pattern Recognition*, pp. 13142–13153, 2023.

[26] B. Li, Y. Zhang, L. Chen, J. Wang, F. Pu, J. Yang, C. Li, and Z. Liu, "Mimic-it: Multi-modal in-context instruction tuning," *arXiv preprint arXiv:2306.05425*, 2023.

[27] S. Abdel-Khalek, A. D. Algarni, G. Amoudi, S. Alkhalaf, F. M. Alhomayani, and S. Kathiresan, "Leveraging ai-generated content for synthetic electronic health record generation with deep learning-based diagnosis model," *IEEE Transactions on Consumer Electronics*, pp. 1–1, 2024.

[28] P. Mishra, M. Warr, and R. Islam, "Tpack in the age of chatgpt and generative ai," *Journal of Digital Learning in Teacher Education*, vol. 39, no. 4, pp. 235–251, 2023.

[29] H. Liu, C. Li, Y. Li, and Y. J. Lee, "Improved baselines with visual instruction tuning," in *Proceedings of the IEEE/CVF Conference on Computer Vision and Pattern Recognition*, pp. 26296–26306, 2024.